\documentclass[twocolumn,preprintnumbers,showpacs,amsmath,amssymb]{revtex4}
\usepackage{graphicx}
\usepackage{amssymb}

\begin{document}
\title{Quantum protocols for the millionaire problem with a third party are
trivial}
\author{Guang Ping He}
\email{hegp@mail.sysu.edu.cn}
\affiliation{School of Physics and
Engineering, Sun Yat-sen University, Guangzhou 510275, China}

\begin{abstract}
Recently there were many quantum protocols devoted to solve the
millionaire problem and private comparison problem by adding a
semi-honest third party. They all require complicated quantum
methods, while still leak a non-trivial amount of information to at
least one of the parties. But it will be shown here that once the
third party is introduced, there are very simple protocols which
require quantum key distribution as the only quantum resource, and
the amount of information leaked can be made arbitrarily small.
Furthermore, even a dishonest third party cannot spoil the
protocols. Thus our solutions surpass all existing protocols on both
feasibility and security.
\end{abstract}



\maketitle

\newpage


\section{Introduction}

The millionaire problem \cite{m1} was originally a two-party secure
computation problem, in which two millionaires, Alice and Bob, want
to know which of them is richer without revealing their actual
wealth. It is analogous to a more general problem whose goal is to
compare two numbers $a$ and $b$, without revealing any extra
information on their values other than what can be inferred from the
comparison result. There is also a variation called the socialist
millionaire problem \cite{m2}, in which Alice and Bob want to
determine if their wealth $a$ and $b$ are equal, without disclosing
any extra information on the values of $a$ and $b$ to each other. As
typical examples of secure multi-party computations, these problems
play essential roles in cryptography. They have many applications in
e-commerce and data mining where people need to compare numbers
which are confidential.

Nevertheless, the original solution \cite{m1} to the problems needs to rely
on oblivious transfer \cite{qbc9}, which is hard to achieve unconditional
security even in quantum cryptography \cite{qi500,qi499,qi677,qi725,qbc14}.
Therefore, people considered a relaxed setting of the problems which
involves an additional semi-honest third party, generally called Trent (or
TP). Trent communicates with Alice and Bob separately. He is regarded as
semi-honest because, on one hand, they study only the case where he executes
the protocol faithfully, and loyally keeps the data he exchanges with one
party secret from the other. That is, he will not try to spoil the protocol,
nor he will help either Alice or Bob to cheat. On the other hand, he is not
fully trustable as he may attempt to learn the values of $a$ and $b$ or the
comparison result, by methods such as eavesdropping or intercepting the
classical and quantum channels between the other two parties, or faking the
quantum states which look authentic to Alice and Bob while entangled with
his ancillary systems that can provide him additional informations, etc.

Under this scenario, the problems become a three-party cryptography, so that
the existing impossibility proofs \cite{qi500,qi499,qi677,qi725,qbc14} on
two-party secure computations do not necessarily apply. Therefore, many
quantum protocols were proposed. Jia et al. \cite{qi1042} gave a solution to
the millionaire problem, while the others \cite%
{qi1050,qi1043,qi1049,qi1120,qi1044,qi1045,qi1047,qi1116,qi1046,qi1048,qi1115,qi1110,qi1117}
studied the socialist millionaire problem under the name \textquotedblleft
quantum private comparison\textquotedblright\ (QPC).

In this paper, however, we will show that once the third party is included,
there exists simple solutions to the problems which are basically classical
protocols with the assistance of quantum key distribution (QKD) \cite{qi365}%
. On the contrary, all previous protocols for the millionaire
problem and QPC require a much greater amount of quantum resources,
such as entanglement, joint measurements, quantum memory, decoy
states, etc. Therefore they are all inferior in feasibility and
simplicity.

Moreover, our protocols have two security advantages over the others. First,
we will show that all previous protocols \cite%
{qi1042,qi1050,qi1043,qi1049,qi1120,qi1044,qi1045,qi1047,qi1116,qi1046,qi1048,qi1115,qi1110,qi1117}
inevitably leak a non-trivial amount of information to Trent or both Alice
and Bob, while in our protocols this can be avoided. Secondly, a dishonest
Trent can spoil many of the previous protocols \cite%
{qi1042,qi1120,qi1044,qi1045,qi1047,qi1116,qi1046,qi1048,qi1115,qi1110,qi1117}%
. That is, he can mislead Alice and Bob to wrong results even though
it brings no advantage to himself. But in our protocol such a
dishonest behavior can be detected with a probability that can be
made arbitrarily close to $100\%$. Thus the only constraint on Trent
in our protocols is that he should not cheat together with either
Alice or Bob to steal the secret data of the other party. Other than
that, Trent can be fully distrustful.

In the next section, we will propose our simple protocol for the
millionaire problem, and prove its security. Then we will show in
section III how to adapt the protocol for the QPC task. A detailed
comparison on the feasibility and security of our proposal and
previous protocols will be provided in section IV.

\section{Simple protocol for the millionaire problem}

Suppose that Alice has a secret number $a$, and Bob has a secret number $b$.
Like all previous works \cite%
{qi1042,qi1050,qi1043,qi1049,qi1120,qi1044,qi1045,qi1047,qi1116,qi1046,qi1048,qi1115,qi1110,qi1117}
on the subject, in this paper we only consider the case where Alice and Bob
will not try to spoil the protocol. That is, they are willing to help each
other to calculate the correct comparison result between $a$ and $b$
honestly, without trying to input a wrong value of the secret number. The
only cheating that we need to take care of is that they may attempt to learn
extra information on the value of the secret number of the other party,
other than what the comparison result naturally implies. Also, the third
party Trent will not help either Alice or Bob to cheat.

When dealing with the original millionaire problem, we also assume that the
case $a=b$ will never happen. Otherwise the comparison result will
inevitably reveal the values of $a$ and $b$\ to both parties, so that it
will be impossible to reach the original goal of the millionaire problem,
which requires the values to remain unrevealed. Note that the previous
protocol \cite{qi1042} did not cover this case either.

\subsection{The basic protocol}

For easy understanding, we first consider the case where Trent will not try
to spoil the protocol. To find out which one of $a$ and $b$ is larger with
the help of Trent, Alice and Bob can use the following protocol.

\bigskip

\textit{Protocol P0:}

(1) Alice and Bob share two random numbers $c$ and $\lambda $ ($\lambda \neq
0$) through QKD. There is no restriction on the selection range of $c$ and $%
\lambda $ as long as they are both real numbers. Their order of
magnitude does not need to match that of $a$ and $b$. They can be
either positive or negative, and $\left\vert \lambda \right\vert $
can be either larger or smaller than $1$. Of course, from a
practical point of view, Alice and Bob can limit $c$ and $\lambda $
to a finite range of rational numbers with a certain precision, so
that they can be determined efficiently through QKD. But it is
important to keep these range and precision secret from Trent, in
order to minimize the information on $a$ and $b$ leaked to Trent.

(2) Alice calculates%
\begin{equation}
\alpha =\lambda a+c,  \label{eq1}
\end{equation}%
and sends $\alpha $ to Trent through QKD.

(3) Bob calculates%
\begin{equation}
\beta =\lambda b+c,  \label{eq2}
\end{equation}%
and sends $\beta $ to Trent through QKD.

(4) Trent calculates%
\begin{equation}
D=\alpha -\beta ,  \label{eq3}
\end{equation}%
and sets $R=0$ ($R=1$) if $D>0$ ($D<0$). Then he announces $R$ to both Alice
and Bob publicly.

(5) Alice and Bob will both know the comparison result from $\lambda $ and $%
R $. If $(-1)^{R}\lambda >0$ then the result is $a>b$, otherwise
$a<b$.

\bigskip

Note that any information (including language, images, etc.) can be
encoded digitally, and then presented as binary strings and
transferred via QKD, so that the parties can virtually
\textquotedblleft talk\textquotedblright\
anything with QKD. Therefore in the above protocol, $a$, $b$, $c$ and $%
\lambda $\ are not limited to bits nor integers, and do not have to be
written directly in binary representations and then mapped into quantum
operations bit by bit like they did in previous protocols \cite%
{qi1050,qi1043,qi1049,qi1120,qi1044,qi1045,qi1047,qi1116,qi1046,qi1048,qi1115,qi1110,qi1117}%
. If Alice and Bob want to compare binary strings bit by bit \cite%
{qi1050,qi1043,qi1049,qi1120,qi1044,qi1045,qi1047,qi1116,qi1046,qi1048,qi1115,qi1110,qi1117}%
, or to compare many pairs of large numbers \cite{qi1042}, they can
simply repeat our protocol many times. Note that if both $a$ and $b$
are single bits, then the comparison result will make Alice and Bob
easily deduce the secret number of each other. But this situation is
inevitable by nature of the millionaire
problem and QPC, and also exists in all previous protocols \cite%
{qi1042,qi1050,qi1043,qi1049,qi1120,qi1044,qi1045,qi1047,qi1116,qi1046,qi1048,qi1115,qi1110,qi1117}%
.

The correctness of the protocol can easily be verified. By combining
equations (\ref{eq1}) - (\ref{eq3}), we have $a-b=D/\lambda $. Since $%
(-1)^{R}$ actually represents the sign of $D$, $(-1)^{R}\lambda $ will have
the same sign as that of $a-b$, and thus indicates which one of $a$ and $b$
is larger.

\subsection{The security proof}

A distinct merit of our protocol is that other than using QKD to protect the
communications between the participants, no more quantum method is involved.
The rest parts of the protocol are completely classical. Therefore, given
that QKD is unconditionally secure, the security proof of the protocol is
simple elementary mathematics.

From Trent's point of view, since the information exchanged between Alice
and Bob is secured by QKD, all the information Trent obtained in the
protocol is merely the values of $\alpha $ and $\beta $, and the fact that
they satisfy the relationship described in equations (\ref{eq1}) and (\ref%
{eq2}). The values of $a$, $b$, $c$ and $\lambda $ are not available
directly to him. Since two equations are insufficient for determining four
unknown variates, Trent will find that there could be infinite solutions for
$a$ and $b$ so that he cannot know which one is larger.

In fact, suppose that there are $u$, $v$, $c_{0}$\ and $\lambda _{0}$\
satisfying%
\begin{equation}
\alpha =\lambda _{0}u+c_{0},
\end{equation}%
and%
\begin{equation}
\beta =\lambda _{0}v+c_{0},
\end{equation}%
then it can be verified that they also satisfy%
\begin{equation}
\alpha =(-\lambda _{0})v+(\alpha +\beta -c_{0}),
\end{equation}%
and%
\begin{equation}
\beta =(-\lambda _{0})u+(\alpha +\beta -c_{0}).
\end{equation}%
That is, no matter Alice's and Bob's choices satisfy $a=u$, $b=v$, $c=c_{0}$%
, $\lambda =\lambda _{0}$, or $a=v$, $b=u$, $c=\alpha +\beta -c_{0}$, $%
\lambda =-\lambda _{0}$, Trent will receive the same $\alpha $ and
$\beta $. Therefore, $a>b$ and $a<b$ will both make sense to Trent
so that he cannot tell which one is the actual comparison result.
Also, as $\left\vert \lambda
\right\vert $ can be either larger or smaller than $1$, the value of $%
D=\alpha -\beta$ will not manifest the order of magnitude of $\left\vert
a-b\right\vert $. The existence of $c$ further prevents Trent from getting
information on $\left\vert a/b\right\vert $ by calculating $\alpha / \beta$.
Thus the protocol is unconditionally secure against Trent.

From Alice's point of view, since the information exchanged between Bob and
Trent is secured by QKD, she cannot know $\beta $. Consequently, besides her
own $a$, all the information she obtained in the protocol is merely the
values of $c$, $\lambda $\ and$\ R$. Here $c$ and $\lambda $\ are randomly
chosen by her and Bob, which contain no information about $b$. Meanwhile, $R$
carries $1$ bit of information only. According to information theory, this
amount is insufficient to determine $b$ as long as the number of possible
values of $b$ is more than $3$. Therefore the protocol is also
unconditionally secure against Alice. The security against Bob can be proven
similarly.

\subsection{The complete protocol}

Now let us deal with the case where Trent tries to spoil the protocol, i.e.,
he wants to mislead Alice and Bob to a wrong result of the comparison. In
the above Protocol P0, this can be done by announcing a wrong value of $R$
in step (4). That is, when Trent finds $R=0$ ($R=1$), he announces $R=1$ ($%
R=0$) instead. Consequently, Alice and Bob will both obtain a wrong
result on the relationship between $a$ and $b$, without knowing that
Trent has
played the trick. Note that most previous protocols \cite%
{qi1042,qi1120,qi1044,qi1045,qi1047,qi1116,qi1046,qi1048,qi1115,qi1110,qi1117}
have the same problem too. But here we can avoid this cheating by extending
Protocol P0 into the following complete protocol for the millionaire problem.

\bigskip

\textit{Protocol P1:}

(i) Let $a$ and $b$ denote Alice's and Bob's secret numbers,
respectively, that they want to compare. They decide and share a
large integer $n$ and an index $i_{0}\in \lbrack 1,n]$ through QKD,
and keep them secret from Trent.

(ii) Alice and Bob choose $n$ pairs of numbers $a_{i}$ and $b_{i}$ ($%
i=1,2,...,n$) in the following way. For $i=i_{0}$, Alice takes $a_{i}=a$ and
Bob takes $b_{i}=b$. For all other $i$'s, they take $a_{i}$ and $b_{i}$ as
two numbers that are known to each other but kept secret from Trent, e.g.,
for simplicity Alice can always takes $a_{i}=2$ and Bob takes $b_{i}=1$.

(iii) For $i=1$ to $n$, Alice and Bob compare $a_{i}$ and $b_{i}$ with the
help of Trent using Protocol P0, each time with a different set of the
parameters $c$ and $\lambda $. That is:

\qquad (iii-1) Alice and Bob share two random numbers $c_{i}$ and $\lambda
_{i}$ ($\lambda _{i}\neq 0$) through QKD.

\qquad (iii-2) Alice calculates%
\begin{equation}
\alpha _{i}=\lambda _{i}a_{i}+c_{i},
\end{equation}%
and sends $\alpha _{i}$ to Trent through QKD.

\qquad (iii-3) Bob calculates%
\begin{equation}
\beta _{i}=\lambda _{i}b_{i}+c_{i},
\end{equation}%
and sends $\beta _{i}$ to Trent through QKD.

\qquad (iii-4) Trent calculates%
\begin{equation}
D_{i}=\alpha _{i}-\beta _{i},
\end{equation}%
and sets $R_{i}=0$ ($R_{i}=1$) if $D_{i}>0$ ($D_{i}<0$). Then he announces $%
R_{i}$ to both Alice and Bob publicly.

(iv) Alice and Bob will both know the comparison result between $a$
and $b$ from $\lambda _{i_{o}}$ and $R_{i_{o}}$, i.e., if
$(-1)^{R_{i_{o}}}\lambda _{i_{o}}>0$ then the result is $a>b$,
otherwise $a<b$.

(v) The security check: For each $i\neq i_{0}$, Alice and Bob check whether
Trent's announced $R_{i}$ always matches their $a_{i}$, $b_{i}$ and $\lambda
_{i}$. For example, if they always take $a_{i}=2$ and $b_{i}=1$ ($\forall
i\neq i_{0}$), then there should always be $(-1)^{R_{i}}\lambda _{i}>0$.
Whenever a mismatched value of $R_{i}$ is found, they know that Trent is
cheating.

\bigskip

We can see that Protocol P1 is simply repeating Protocol P0 for $n$ runs,
with each run comparing a different pair of $a_{i}$ and $b_{i}$. In the $%
i_{0}$-th run they compare the values of $a$ and $b$ that they are
really interested, while in the other $n-1$ runs both $a_{i}$ and
$b_{i}$ are known
to them. Therefore, among all the $R_{i}$'s ($i=1,...,n$) Trent announced, $%
n-1$ of them will be checked in step (v). If Trent tries to spoil the
protocol by announcing wrong values of $R_{i}$'s in more than one run, he
will be caught in step (v) with probability $100\%$. On the other hand,
since Trent does not know $i_{0}$, if he announces a wrong value of $R_{i}$
in merely one single run of Protocol P0, there will only be a probability $%
1/n$ that the run he chooses is exactly the $i_{0}$-th run. As a
consequence, by increasing $n$, the probability for Trent to spoil
the protocol without being detected can be made arbitrarily small.

\section{The quantum private comparison protocol}

When applying Protocol P0 directly for the socialist millionaire problem,
a.k.a. quantum private comparison\ (QPC), there will be a security loophole.
That is, Trent can always know the comparison result ($a=b$ or $a\neq b$) by
checking whether there is $D=0$. Note that this is also the case in the
protocols proposed in Refs. \cite{qi1120,qi1044,qi1045,qi1047,qi1116,qi1115,qi1110}%
. But it is surely better if the loophole can be avoided. Here we show that
this goal can indeed be achieved by slightly modifying our above Protocol
P1, as described below.

\bigskip

\textit{Protocol P2:}

(I) Let $a$ and $b$ denote Alice's and Bob's secret numbers,
respectively, that they want to compare. They decide and share two
large integers $m$, $n$ ($m<n$) and a set of indices
$S=\{i_{0},i_{1},i_{2},...,i_{m}\}$ through QKD, and keep them
secret from Trent. Here, each $i_{j}$\ in $S$ is a randomly chosen
integer within the range $[1,n]$, and all $i_{j}$'s have
different values. Note that the values of $i_{0}$, $i_{1}$, $i_{2}$, $...$, $%
i_{m}$\ do not need to be in ascending nor descending order.

(II) Alice and Bob choose $n$ pairs of numbers $a_{i}$ and $b_{i}$ ($%
i=1,2,...,n$) in the following way. If $i=i_{0}$, then Alice takes $a_{i}=a$
and Bob takes $b_{i}=b$. Otherwise Bob always takes $b_{i}=1$, while Alice
takes $a_{i}=1$ if $i\in \{i_{1},i_{2},...,i_{m}\}$, or $a_{i}\neq 1$ if $%
i\notin S$.

(III) For $i=1$ to $n$, Alice and Bob compare $a_{i}$ and $b_{i}$ with the
help of Trent using a process similar to Protocol P0. That is:

\qquad (III-1) Alice and Bob share two random numbers $c_{i}$ and $\lambda
_{i}$ ($\lambda _{i}\neq 0$) through QKD.

\qquad (III-2) Alice calculates%
\begin{equation}
\alpha _{i}=\lambda _{i}a_{i}+c_{i},
\end{equation}%
and sends $\alpha _{i}$ to Trent through QKD.

\qquad (III-3) Bob calculates%
\begin{equation}
\beta _{i}=\lambda _{i}b_{i}+c_{i},
\end{equation}%
and sends $\beta _{i}$ to Trent through QKD.

\qquad (III-4) Trent announces publicly to Alice and Bob whether $\alpha
_{i}=\beta _{i}$ or not.

(IV) Alice and Bob will both know whether $a=b$ or not according to Trent's
announced result on $\alpha _{i_{0}}$ and $\beta _{i_{0}}$ in the $i_{0}$-th
run of step (III-4).

(V) The security check: For each $i\neq i_{0}$, Alice and Bob check whether
Trent's announced result always matches their $a_{i}$, $b_{i}$. That is,
Trent's announcement should always be $\alpha _{i}=\beta _{i}$ for $\forall
i\in \{i_{1},i_{2},...,i_{m}\}$, or $\alpha _{i}\neq \beta _{i}$ for $%
\forall i\notin S$. Otherwise he is cheating.

\bigskip

The key idea of this protocol is: Alice and Bob compare many pairs of $a_{i}$
and $b_{i}$, most of which (except $a_{i_{0}}$ and $b_{i_{0}}$) have nothing
to do with their actual secret numbers $a$ and $b$. The purpose of
introducing these extra pairs is merely to confuse Trent.

The security of the protocol is also obvious. Note that unlike step (4) of
the original Protocol P0, here in step (III-4) Trent merely announces
whether $\alpha _{i}$ and $\beta _{i}$ are equal or not, without announcing
which one is larger. Therefore in the case $a\neq b$, Alice and Bob cannot
deduce whether $a>b$ or $a<b$ like they did in step (5) of P0, so that no
further information on $a$ and $b$ is leaked to them. From Trent's point of
view, no matter $a=b$ or $a\neq b$, he will find $\alpha _{i}=\beta _{i}$
(i.e., $a_{i}=b_{i}$) in some runs of step (III-4). He has no idea whether
these runs include $a$ and $b$ or not, as the values of $i_{0}$ and $m$ are
protected by the QKD process between Alice and Bob. Given that QKD is
unconditionally secure, we achieve the goal that the final comparison
between $a$ and $b$ is kept secret from Trent. Again, no other quantum
methods are required besides QKD.

From the similarity between Protocols P1 and P2, we can easily see
that P2 can also prevent Trent from spoiling the protocol for the
same reason in section II C.

\section{Comparison between existing protocols}

\subsection{Feasibility comparison}

We summarized the comparison on the technical requirement of our
above proposal and previous protocols in Table I. Note that only the
one in Ref. \cite{qi1042} and our Protocol P1 deal with the original
millionaire problem, i.e., finding the larger one among $a$ and $b$.
The rest (including our Protocol P2) are all QPC protocols, which
only compare whether $a$ and $b$ (or $X$ and $Y$ in Refs.
\cite{qi1050,qi1043,qi1049,qi1120,qi1044,qi1045,qi1046,qi1048,qi1115,qi1110},
$S_{A}$ and $S_{B}$ in Ref. \cite{qi1117}) are equal, without
judging which one is larger. Also, Ref. \cite{qi1049} pointed out
that the original protocol in Refs. \cite{qi1050,qi1043} is
insecure, and proposed a corresponding solution. Thus we treated
Refs. \cite{qi1050,qi1043,qi1049} as one protocol in Table I.
Similarly, Ref. \cite{qi1045} pointed out the security loopholes of
the protocol in Ref. \cite{qi1044}, and suggested two improvements,
with one of them making use of the decoy state method. Thus we
treated Refs. \cite{qi1044,qi1045} as one protocol in Table I, and
listed \textquotedblleft decoy states\textquotedblright\ as
\textquotedblleft optional\textquotedblright . On the other hand,
Ref. \cite{qi1116} commented on the security problem of the protocol
in Ref. \cite{qi1047}, and proposed a modification which no longer
requires decoy states as Ref. \cite{qi1047} did. Thus we listed them
separately in the table.

\begin{table*}
\caption{Comparison on the technical requirement of our protocols
P1, P2 and existing millionaire
problem and QPC protocols \protect\cite%
{qi1042,qi1050,qi1043,qi1049,qi1120,qi1044,qi1045,qi1047,qi1116,qi1046,qi1048,qi1115,qi1110,qi1117}.
All the blank spaces mean \textquotedblleft no\textquotedblright .}
\begin{ruledtabular}
\begin{tabular}{lccccc}
&entanglement&joint&quantum&decoy&QKD \\
&&measurements&memory&states& \\ \hline
Ref. \cite{qi1042}'s&Yes&Yes&Yes&Yes&Yes \\
Refs. \cite{qi1050,qi1043,qi1049}'s&Yes&Yes&Yes&Yes&Yes \\
Ref. \cite{qi1120}'s& & &Yes& & \\
Refs. \cite{qi1044,qi1045}'s&Yes& &Yes&optional& \\
Ref. \cite{qi1047}'s&Yes& &Yes&Yes& \\
Ref. \cite{qi1116}'s&Yes& &Yes& & \\
Ref. \cite{qi1046}'s&Yes& &Yes& &Yes \\
Ref. \cite{qi1048}'s&Yes&Yes&Yes& &Yes \\
Ref. \cite{qi1115}'s&Yes&Yes&Yes&Yes&Yes \\
Ref. \cite{qi1110}'s& & &Yes& &Yes \\
Ref. \cite{qi1117}'s& & &Yes& &Yes \\
Ours& & & & &Yes \\
\end{tabular}
\end{ruledtabular}
\end{table*}

As we can see, all previous protocols \cite%
{qi1042,qi1050,qi1043,qi1049,qi1120,qi1044,qi1045,qi1047,qi1116,qi1046,qi1048,qi1115,qi1110,qi1117}
require much more quantum resources than ours. First, quantum memory (at
least short-term one) is required in all these protocols, as there are
always some parts of the quantum states which cannot be measured immediately
once they are received, because the participants cannot determine the
measurement basis or the location of the states used for the security
checks, until they receive some necessary announcement from other parties.
Secondly, most of them (except those in Refs. \cite{qi1120,qi1110,qi1117}) have to rely on quantum entanglement. To compare two numbers $a$ and $b$ satisfying $1<a,b<N$, the protocol in Ref. \cite{qi1042} even requires the use of $2N$-level entangled states. Thirdly, some protocols \cite%
{qi1042,qi1050,qi1043,qi1049,qi1048,qi1115} require the use of joint
measurements on multi-particle states. All these technical requirements
seriously lower the feasibility of the protocols. Furthermore, the decoy
state method is sometimes adopted \cite{qi1042,qi1050,qi1043,qi1049,qi1045,qi1047,qi1115}, which requires a large amount of quantum transmission and thus reduces the efficiency of the
protocols. Even so, some proposals \cite%
{qi1042,qi1050,qi1043,qi1049,qi1046,qi1048,qi1115,qi1110,qi1117} still
involve QKD as parts of the protocols.

On the contrary, in our protocols, other than using QKD to transmit
classical information, no more quantum states and operations are required.
Since there exists QKD protocol \cite{qi365} in which entanglement, joint
measurements and quantum memory are not necessary, our protocols are much
easier to be implemented than all previous proposals.

\subsection{Security comparison}

Table II shows the comparison on the security of all protocols. As
we elaborated in the above sections, our Protocols P1 and P2 manage
to meet the security requirements of the original millionaire
problem and QPC, respectively. Alice and Bob know nothing about the
secret data of the other party, except what can be inferred from the
comparison result. Trent does not know the comparison result at all.
Both P1 and P2 can also prevent a dishonest Trent from spoiling the
protocol.

On the contrary, spoiling the protocol can be done in many previous
proposals \cite%
{qi1042,qi1120,qi1044,qi1045,qi1047,qi1116,qi1046,qi1048,qi1115,qi1110,qi1117}%
, because Trent can simply lie in the final stage. But\ more importantly,
even though all previous protocols \cite%
{qi1042,qi1050,qi1043,qi1049,qi1120,qi1044,qi1045,qi1047,qi1116,qi1046,qi1048,qi1115,qi1110,qi1117}
can prevent Trent from knowing the exact values of $a$ and $b$, they still
leak extra information to either Trent or Alice and Bob. In the protocol for
the millionaire problem in Ref. \cite{qi1042}, Trent always knows $\left\vert
a-b\right\vert $, while the QPC protocols in Refs. \cite%
{qi1120,qi1044,qi1045,qi1047,qi1116,qi1115,qi1110} have the problem that
Trent knows whether $a=b$ or not, as it was clearly shown in the final steps
of these protocols. The rest QPC protocols \cite%
{qi1050,qi1043,qi1049,qi1046,qi1048,qi1117} is secure against Trent, but
besides the comparison result ($a=b$ or $a\neq b$), Alice and Bob will still
obtain an extra amount of information on $a$ and $b$, as elaborated below.

\begin{table*}
\caption{Comparison on the security of our protocols P1, P2 and
existing millionaire problem and QPC protocols \protect\cite%
{qi1042,qi1050,qi1043,qi1049,qi1120,qi1044,qi1045,qi1047,qi1116,qi1046,qi1048,qi1115,qi1110,qi1117}.
All the blank spaces mean \textquotedblleft no\textquotedblright .}
\begin{ruledtabular}
\begin{tabular}{lccc}
&Alice/Bob gains&Trent knows&Trent can spoil \\
&extra information&the result&the protocol \\ \hline
Ref. \cite{qi1042}'s& &partial&Yes \\
Refs. \cite{qi1050,qi1043,qi1049}'s&Yes& & \\
Ref. \cite{qi1120}'s& &Yes&Yes \\
Refs. \cite{qi1044,qi1045}'s& &Yes&Yes \\
Ref. \cite{qi1047}'s& &Yes&Yes \\
Ref. \cite{qi1116}'s& &Yes&Yes \\
Ref. \cite{qi1046}'s&Yes& &Yes \\
Ref. \cite{qi1048}'s&Yes& &Yes \\
Ref. \cite{qi1115}'s& &Yes&Yes \\
Ref. \cite{qi1110}'s& &Yes&Yes \\
Ref. \cite{qi1117}'s&Yes& &Yes \\
Ours& & & \\
\end{tabular}
\end{ruledtabular}
\end{table*}

In the QPC protocol in Refs. \cite{qi1050,qi1043,qi1049},\ Alice and Bob
first express $a$ and $b$ in binary representations as $%
a=a^{(1)}a^{(2)}...a^{(i)}...$ and $b=b^{(1)}b^{(2)}...b^{(i)}...$ ($%
a^{(i)},b^{(i)}\in \{0,1\}$ for all $i$'s), respectively, then they randomly
pick $i=i_{1},i_{2},...$ and compare each pair of the bits $a^{(i)}$ and $%
b^{(i)}$ one by one with the help of Trent. If all pairs of $a^{(i)}$ and $%
b^{(i)}$ turn out to be equal then they know that $a=b$. But once they find
a difference pair (e.g., $a^{(i_{d})}\neq b^{(i_{d})}$) which indicates that
$a\neq b$, they should immediately abort the procedure without further
comparing the rest $a^{(i)}$'s and $b^{(i)}$'s. In this case, only the first
few compared bits of $a$ and $b$ are known to be $a^{(i)}=b^{(i)}$ ($%
i=i_{1},i_{2},...,i_{d-1}$) and $a^{(i_{d})}\neq b^{(i_{d})}$. This
is insufficient to determine the exact values of $a$ and $b$.
However, Bob (Alice) will know the exact values of the $d$ bits of
$a$ ($b$) that are already compared, i.e., they both gain $d$\
($d\geq 1$) bits of information about $a$ and $b$. Since $d>1$
occurs with a non-vanishing probability, the average amount of
information gained will be larger than, and unequal to $1$ bit. Note
that in Refs. \cite{qi1050,qi1043,qi1049},\ Alice's and Bob's actual
secret numbers that they want to compare are $X$ and $Y$,
respectively, while $a=H(X)$ and $b=H(Y)$ are their corresponding
hash values, where $H$ is a secret hash function they share
beforehand. Nevertheless,
a good hash function useful for QPC has to be a $1$-to-$1$ mapping between $X$ and $a$ ($%
Y$ and $b$), otherwise there could be the case where $X\neq Y$ while the QPC
protocol outputs $a=b$. Therefore, knowing $d$\ bits of $a$ ($b$) means that
the possible choices of the value of $X$ ($Y$) will be limited to those
whose hash values contain the bits $a^{(i)}$ ($b^{(i)}$), $%
i=i_{1},i_{2},...,i_{d}$. That is, there are also $d$\ bits of mutual
information about $X$ ($Y$) which become known to Bob (Alice). The use of
the hash function merely changes the type of the information leaked, while
the amount of this information remains the same.

Similarly, in the protocol in Ref. \cite{qi1117} Alice and Bob also
compare the hash values of their secret numbers bit by bit.
Therefore, for the same reason, when they know that their secret
numbers are unequal, there were already many bits of the hash values
become known to both of them, so that a nontrivial amount of mutual
information is leaked.

The protocols in Refs. \cite{qi1046,qi1048} leak extra information
to Alice and Bob too. After calculating $R$ in equation (15) of Ref.
\cite{qi1046} or equation (9) of Ref. \cite{qi1048}, if $R\neq 0$,
Alice and Bob will not only know that $X\neq Y$, but also know that
the number of different bits in the binary representation of $X$ and
$Y$ is exactly $R$. Therefore, the number of the possible choices of
the value of $Y$ ($X$) will be limited to $\left(
\begin{array}{c}
N \\
R%
\end{array}%
\right) $, where $N$ is the length of the binary representation of $X$ and $%
Y $. Before the comparison, the number of choices was $2^{N}$. Thus
the amount of mutual information leaked to each of Alice and Bob is
$I(R)=\log _{2}2^{N}-\log _{2}\left(
\begin{array}{c}
N \\
R%
\end{array}%
\right) =N-\log _{2}\left(
\begin{array}{c}
N \\
R%
\end{array}%
\right) $. If all values of $X$ and $Y$ occur with equal probabilities, the
probability for them to have $R$ different bits will be $prob(R)=\left(
\begin{array}{c}
N \\
R%
\end{array}%
\right) /2^{N}$. Considering that $X\neq Y$\ happens with the probability $%
p=(2^{N}-1)/2^{N}$, when it indeed happened, the average amount of
mutual information leaked to each of Alice and Bob will be $\bar{I}%
=(1/p)\sum\nolimits_{R=1}^{N}prob(R)\cdot
I(R)=N-[\sum\nolimits_{R=1}^{N}\left(
\begin{array}{c}
N \\
R%
\end{array}%
\right) \log _{2}\left(
\begin{array}{c}
N \\
R%
\end{array}%
\right) ]/(2^{N}-1)>1$. That is, knowing the value of $R$ makes Alice and
Bob each gain more than $1$ bit of extra information from these protocols.

Thus we see that all previous protocols \cite%
{qi1042,qi1050,qi1043,qi1049,qi1120,qi1044,qi1045,qi1047,qi1116,qi1046,qi1048,qi1115,qi1110,qi1117}
leak extra information to the parties. Our P1 and P2 win hands-down
in this category as there is no information leaked.

\section{Summary}

Thus we show that with the presence of a third party, the
millionaire problem and QPC can be solved with protocols which
require QKD as the only quantum resource. There is no need for
entanglement, joint measurements, quantum memory, etc.. As QKD is
already well-developed experimentally, our protocols are fully
feasible with currently available technology.

Moreover, all previous protocols leak extra informations to at least one of
the party. Most of them can also be spoiled by a dishonest third party. Our
protocols manage to fix all these security problems.


The work was supported in part by the NSF of China under grant No.
10975198, the NSF of Guangdong province, and the Foundation of
Zhongshan University Advanced Research Center.

\end{document}